# Thermodynamics of a bad metal–Mott insulator transition in the presence of frustration

J. Kokalj[*] and Ross H. McKenzie

*School of Mathematics and Physics, University of Queensland, Brisbane, 4072 Queensland, Australia*

(Dated: April 26, 2013)

Thermodynamic properties of the Hubbard model on the anisotropic triangular lattice at half filling are calculated by the finite-temperature Lanczos method. The charge susceptibility exhibits clear signatures of a metal-Mott insulator transition. The metallic phase is characterized by a small charge susceptibility, large entropy, large renormalized quasiparticle mass, and large spin susceptibility. The fluctuating local magnetic moment in the metallic phase is large and comparable to that in the insulating phase. These bad metallic characteristics occur above a relatively low coherence temperature, as seen in organic charge transfer salts.

Remarkable observations of a possible spin liquid phase [1] and a new universality class of the metal-insulator transition [2] in organic charge transfer salts, which in addition show unconventional superconductivity [3], have increased interest in these materials. It has been argued that a proper microscopic description of these material can be given with a Hubbard model on the anisotropic triangular lattice at half filling [4]. Parameters of the model for the description of organic charge transfer salts fall into the regime of strong correlations and significant frustration of antiferromagnetic spin interactions. This is the most challenging parameter regime, where analytical approaches become unreliable, and one needs to resort to numerical techniques.

In this Letter we study a range of thermodynamic properties (charge susceptibility, specific heat, entropy and spin susceptibility) of the Hubbard model on the anisotropic triangular lattice at half-filling. The model exhibits a Mott metal-insulator transition (MIT), which can be driven either by interaction strength or by frustration. We argue that the metallic phase has a strongly reduced coherence temperature $T_{\rm coh}$, below which a Fermi liquid metal with coherent quasiparticle excitations may exist. Above $T_{\rm coh}$ the model is in a bad metallic regime with large local magnetic moments. We show how frustration increases the low temperature specific heat, entropy and spin susceptibility in the insulating phase. Although the charge susceptibility shows definitive signatures of the metal-insulator transition, the specific heat and spin susceptibility do not. Indeed, above $T_{\rm coh}$ there appears to be little difference between the bad metal and the Mott insulator. This is similar to the dynamical mean-field theory (DMFT) picture of the transition [5, 6].

*Model.* The Hubbard model on the anisotropic triangular lattice has the Hamiltonian

$$H = -\sum_{i,j,\sigma} t_{ij} c^\dagger_{i,\sigma} c_{j,\sigma} + U \sum_i n_{i\uparrow} n_{i\downarrow} - \mu \sum_{i,\sigma} n_{i,\sigma}. \quad (1)$$

The hopping parameters $t_{ij} = t$ for nearest neighbors in two directions of the triangular lattice, while $t_{ij} = t'$ for nearest neighbors in the third direction. $c_{i,\sigma}$ ($c^\dagger_{i,\sigma}$) is a fermionic annihilation (creation) operator for an electron on site $i$ with spin $\sigma$ (either $\uparrow$ or $\downarrow$). $n_{i,\sigma} = c^\dagger_{i,\sigma} c_{i,\sigma}$, $U$ is the on-site Coulomb repulsion, and $\mu$ is the chemical potential. Most of our results are presented in units of $t$, and we use $\hbar = k_B = 1$. We only consider the case of half-filling since this is relevant to several important families of organic charge transfer salts [4].

*Numerical method.* To calculate thermodynamic properties for the model we use the finite-temperature Lanczos method (FTLM) [7–9]. Within FTLM, the Hamiltonian is effectively diagonalized on a small cluster. We use 16 site clusters with twisted boundary conditions. More details on the calculation of thermodynamic properties with FTLM can be found in the Supplementary Material [10] and Refs. 8 and 11. Recently, it was shown that for the *t-J* model, the FTLM gives results in agreement with those obtained by a numerical linked-cluster algorithm [12] suggesting that FTLM on small lattices can give results comparable to the thermodynamic limit for strongly correlated metallic phases.

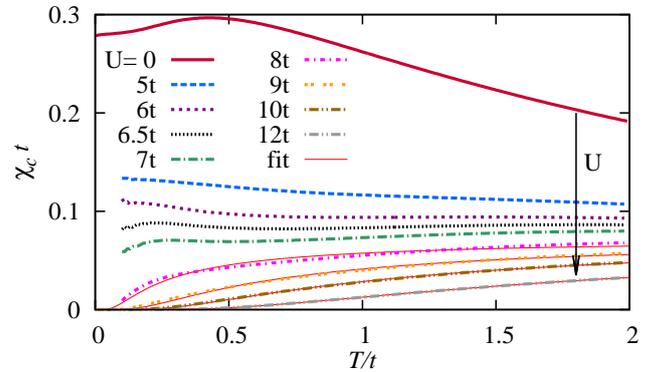

Figure 1. (color online) Signatures of the Mott transition in the temperature dependence of the charge susceptibility $\chi_c$. The figure shows $\chi_c$ vs. temperature $T$ for several interaction strengths $U$ and $t' = t$. $\chi_c$ decreases with increasing $U$ and is almost independent of temperature in the metallic phase ($U \leq 7t$). In contrast, in the insulating phase ($U \geq 8t$) it is strongly suppressed at low $T$, with an activated behavior $\chi_c^{(i)} = a e^{-\Delta_c/T}$ (fits shown with thin red lines).

*Charge susceptibility.* In Fig. 1 we show the temperature dependence of the charge susceptibility $\chi_c \equiv \frac{\partial n}{\partial \mu}$, which is strongly suppressed with increasing $U$ from its non-interacting electron value (calculated for infinite system). This is primarily due to broadening of the density of states over a larger energy range ($W + U$) or due to reduced quasiparticle weight [10]. In addition to this overall decrease of $\chi_c$ with increasing $U$, $\chi_c$ becomes further suppressed at low $T$ for higher $U > U_c$ due to the MIT and opening of a charge gap $\Delta_c$ (see Fig. 2). In the insulating phase, $\chi_c$ shows an activated behaviour $\chi_c^{(i)} = ae^{-\Delta_c/T}$, which allows us to extract $\Delta_c$ from the $T$ dependence of $\chi_c$. Supplementary material [10] shows a plot of $\ln(\chi_c)$ vs. $1/T$. The simultaneous strong decrease of the low-$T$ $\chi_c$ and the opening of $\Delta_c$ with increasing $U$ allows us to extract the critical value of the interaction $U_c$ at which the MIT appears.

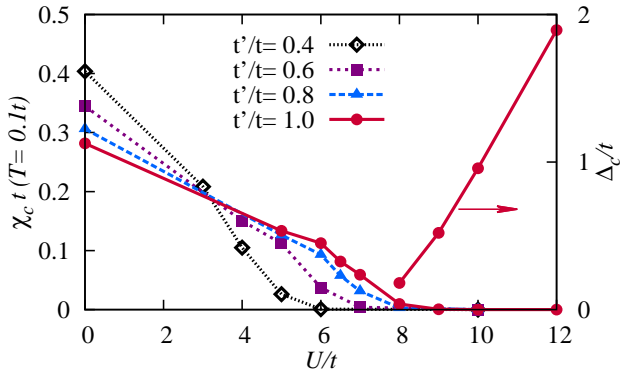

Figure 2. (color online) Signatures of the Mott transition in the charge susceptibility $\chi_c$ at $T = 0.1t$ and charge gap $\Delta_c$. Figure shows the reduction in $\chi_c(T = 0.1t)$ with increasing $U$ for several frustrations $t'/t$ and the charge gap $\Delta_c$ vs. $U$ for $t' = t$. $\chi_c(T = 0.1t)$ is finite and slowly decreasing with increasing $U$ at smaller values of $U$ ($< U_c$), which corresponds to metallic behavior. At some larger value of $U \sim U_c$, $\chi_c(T = 0.1t)$ becomes strongly suppressed exhibiting a metal-insulator transition (MIT) and is close to zero for larger $U$ ($> U_c$), corresponding to Mott insulating behavior. In this regime a charge gap opens, which increases with increasing $U$. The figure also shows that the MIT appears at smaller $U_c$ for less frustrated systems (smaller $t'/t$). This is observed also with the move of $\Delta_c$ curves to the left with decreasing $t'/t$ (not shown).

In Fig. 2 we see that $\chi_c$ does not exhibit any sign of divergence for $U \to U_c$ in the metallic regime ($U < U_c$). This is in contrast to what was observed for a filling-controlled MIT within DMFT [13] and a path-integral renormalization group approach on the square lattice with next-nearest-neighbor hopping [14, 15]. On the other hand, no sign of divergence was observed for the filling controlled MIT in an exact diagonalization study on the triangular lattice [16]. This suggests that upon changing from a filling- to bandwidth-controlled transition, dimensionality or frustration can affect the type of Mott MIT. The possibility of different characters of filling- and bandwidth-controlled MIT was pointed out in Ref. 14.

Although our results at finite $T$ do not allow precise determination of the order of the MIT, the linear dependence of $\Delta_c$ on $U$ (Fig. 2), which persists down to $U$ quite close to $U_c$, is in agreement with a V- or $\Upsilon$-shaped metal-insulator boundary in the $\mu$-$U$ plane and therefore also in agreement with the suggested [14, 17] first order transition. However, our results cannot rule out a second order phase transition, as proposed by Senthil [18]. In the critical regime one expects $\chi_c \propto T$ and close to the critical regime $\chi_c(T = 0) \propto (U_c - U)^\nu$ and $\Delta_c \propto (U - U_c)^\nu$ with $\nu = 0.67$ [18]. Our results in Figs. 1 and 2 do not show signatures of such behaviour, suggesting that the critical region is quite narrow in $T$ and $U$.

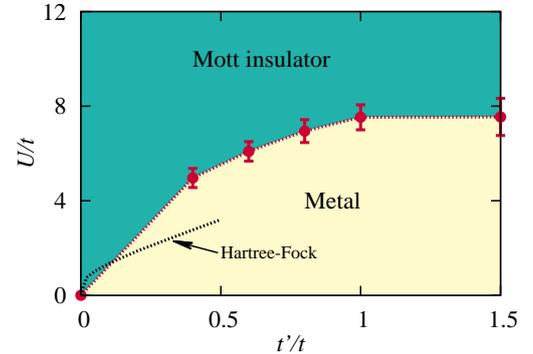

Figure 3. (color online) Zero-temperature phase diagram in the $U$-$t'$ plane. Red points with error bars show our estimate of the Mott metal-insulator transition ($U_c$) for various values of frustration $t'$. Above $U_c$ is a Mott insulating phase and below $U_c$ is a metallic phase. Transition at $t'/t = 0$ corresponds to a square lattice with perfect nesting and appears at $U_c = 0$ [19, 20]. For small $t'/t$, a superlinear increase of $U_c(t')$ is predicted by the Hartree-Fock approximation (shown with black dotted line). The error bars shown were estimated from the range of $U$ values in which $\chi_c(T = 0.1t) < 0.06t$ and $\Delta_c < 0.2t$.

*Phase diagram.* In Fig. 3 we show our estimate of the critical interaction strength $U_c$ (for various $t'/t$) at which the system undergoes a Mott metal-insulator transition. $U_c$ decreases with decreasing frustration $t'/t$ and the MIT can therefore be driven either with increasing interaction strength $U$ or decreasing frustration ($t'/t$). At $t' = 0$ the model is a nearest-neighbour square lattice Hubbard model with perfect nesting for which $U_c = 0$ [10, 19–21] and going away from perfect nesting with increasing $t'/t$ results within a Hartree-Fock approximation in a superlinear increase of $U_c$ (see Fig. 3).

Our phase diagram is consistent with previous findings by several numerical and analytical techniques (see Supp. [10] and Table I therein for more details).

The calculated $T$-dependence of thermodynamic quantities do not show strong signatures of possible different Mott insulator spin states (antiferromagnetic order for small $t'/t$ [22–25], 120 degree Néel order at higher $U$ [23, 26, 27] for $t'/t \sim 1$ and a possible spin liquid at $U \gtrsim U_c$ for $t'/t \sim 1$ [24–27]). The spin structure factor or discontinuities in the double occupancy [10, 27] would be better indicators [28]. Hence, we don't show possible spin ground states in the phase diagram.

*Specific heat and entropy.* In Fig. 4 we show how the $T$-dependence of the specific heat $C_V$ and entropy per site $s$ change with increasing $U$. In the metallic regime ($U < U_c$), the low-$T$ slope of $C_V$ vs. $T$ and $s$ increases with $U$. In a Fermi liquid picture this slope increase corresponds to the increased renormalized quasiparticle mass, and we estimate it reaches $m^*/m_b \sim 2.5$ for $U$ close to $U_c$. $m^*$ is the renormalized quasiparticle mass and $m_b$ is the bare band mass. Comparable enhancements are seen in organic charge transfer salts [29]. Simultaneously with increasing slope, the low-$T$ peak in $C_V$ moves to lower $T$, resulting in a decreased coherence temperature $T_{\text{coh}}$ with increasing U.

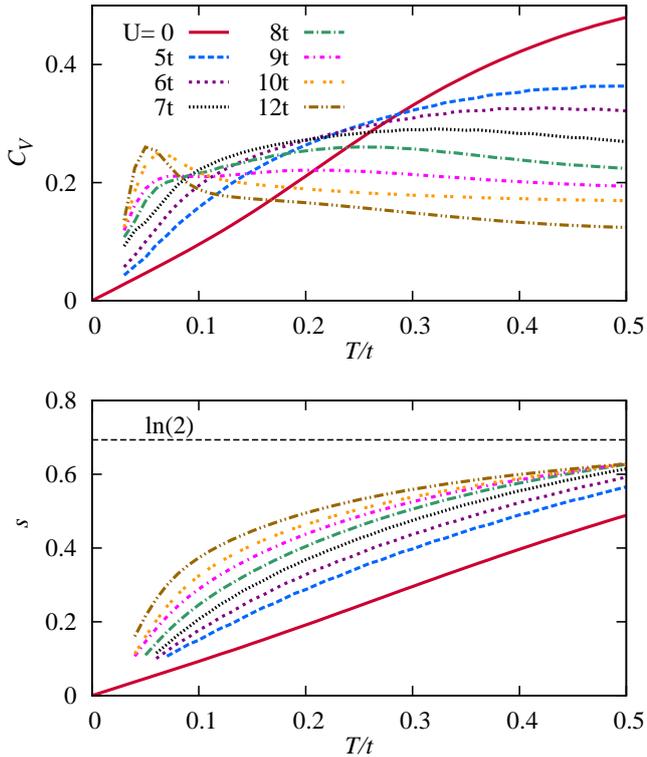

Figure 4. (color online) Temperature dependence of the specific heat $C_V$ (top) and entropy $s$ (bottom) for several values of the interaction $U$ and for $t' = t$. Top: With increasing $U$, $C_V$ increases at low temperatures. For $U < U_c$ this increase is due to an increased quasiparticle renormalized mass manifested also in an increased slope of $C_V$ at low $T$ and a decreased coherence temperature $T_{\text{coh}}$. For $U > U_c$, $C_V$ starts to develop a peak at low $T$ due to well-defined low energy spin excitations, while charge excitations are gapped with a large charge gap $\Delta_c$, resulting in a high $T$ peak (not shown here, see Fig. S7 in Supp. [10]). Bottom: A similar increase is seen in the entropy, which for $U > U_c$ starts to approach the value of $\ln(2)$. This is characteristic of the development of $S = 1/2$ local moments. In particular the strong increase of the entropy at low $T (\simeq 0.1t)$ and plateauing below $\ln(2)$ for $T > 0.3t$ is in contrast to what is observed for $U = 0$: the entropy steadily increases with $T$ and only tends towards $\ln(2)$ at much higher temperatures. Entropy is increased at low $T$ even for $U \lesssim U_c$, signaling the development of local moments already in the metallic regime for $U \sim U_c$, and therefore the bad metallic behavior.

We estimate the coherence temperature $T_{\text{coh}}$ as the temperature at which $C_V$ starts to deviate substantially from linearity in $T$ and obtain $T_{\text{coh}} < 0.1t$ for $t' = t$ in the vicinity of the MIT ($U \sim U_c$). This shows the importance of strong correlations, since $T_{\text{coh}}$ is much smaller than the estimate of $T_{\text{coh}} \sim 0.4t$ for $U = 0$. Electronic structure calculations based on density-functional theory (DFT) give values of $t$ in the range 50-70 meV for the $\kappa$-(BEDT-TTF)$_2$X family [30–32] and 40-50 meV for the $\beta$'-X[Pd(dmit)$_2$]$_2$ family [33]. The ratio of $t'/t$ varies between about 0.4 and 1.3 depending on the counterion X. Taking $t \sim 40$ meV we estimate $T_{\text{coh}} < 50$ K, which is in good agreement with experiments [4]. This temperature corresponds to the vanishing of the Drude peak in the optical conductivity [6], maximum in the thermopower vs. temperature [34] or the resistivity becoming comparable to the Mott-Ioffe-Regel limit [35].

At $T > T_{\text{coh}}$ we expect bad metallic behaviour with well formed local moments. This is supported by the entropy showing already in the metallic regime an increase towards the large $U$ result, and furthermore by the large spin susceptibility close to the result for a Heisenberg model (see Fig. 5). With further increase of $U$ and entering into the insulating regime ($U > U_c$), both $C_V$ and $s$ are strongly increased at low $T$ due to well formed local moments and a large density of low lying spin excitations. This is a hallmark of magnetic frustration [36]. Closer examination of $C_V$ reveals that the low-$T$ peak (at $T < 0.1t$) is strongly increased for $U = 10t$ and $12t$, which might be a signature of a transition from a spin liquid state into a Néel order state, in agreement with the findings in Ref. [27]. High-$T$ properties of $C_V$ and its two peak structure at large $U$ (the low-$T$ peak is due to spin excitations and the high-$T$ peak due to charge excitations) are shown in Fig. S7 of the Supplementary material [10], and for $t' = 0$ were previously discussed in Ref. [11].

*Spin susceptibility.* In Fig. 5 we show, that the spin susceptibility $\chi_s$ is close to the Curie-Weiss (CW) result [36] even in the metallic phase. This gives strong support that already for $U \sim U_c$ the local moment is well formed, resulting in increased $\chi_s$ and a bad metal behaviour for $U \lesssim U_c$.

Agreement with the CW behaviour supports the picture of a bad metal with short range antiferromagnetic correlations, while longer range correlations may suppress $\chi_s$ below CW at quite low $T$. Correlations start to develop below $0.5t$ and become stronger for $T < 0.1t$. Alternatively, the suppression of $\chi_s$ for $T < 0.1t$ might occur due to a spin gap to triplet excitations. Such suppression of longer range spin correlations is due to magnetic frustration [36].

Even in the metallic phase we observe a suppressed $\chi_s$ at low $T$ (see Fig. S10 in Supp. [10]). A similar suppression has been observed in the $T$ dependence of the NMR Knight shift in some organic charge transfer salts [37]. It has been argued that this suppression together with a similar suppression in the NMR relaxation time $1/T_1T$ [37–39] is a signature of a pseudogap. We further discuss this issue in the Supp. [10].

A related subtle and important question concerns the fate of the local moments when $T$ is lowered from the bad metallic to the Fermi liquid regime. From the point of view of DMFT, when the $T$ is lowered below $T_{\text{coh}}$ the quasiparticles form and begin to screen the local moments in the sense of Kondo [5]. However, in the actual system this Kondo screening is competing with the nearest-neighbour antiferromagnetic interactions between the local moments. In the metallic phase the Kondo screening must be dominant. Hence, at the lowest $T$,

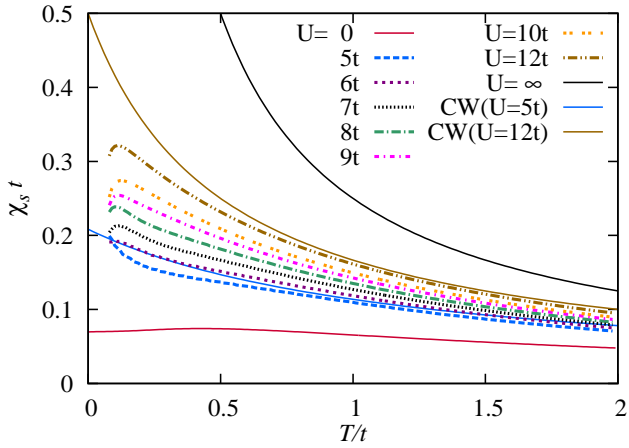

Figure 5. (color online) Spin susceptibility $\chi_s$ vs. $T$ for several values of $U$ and $t'/t = 1$. $\chi_s$ increases with increasing $U$ towards the Curie law $\chi_s(U = \infty) = 1/(4T)$ or noninteracting spin 1/2 result. For intermediate $U$, $\chi_s$ is close to the Curie-Weiss (CW) law $\chi_s^{(CW)} = 1/4(T + T_{CW})$ with $T_{CW} = J + J'/2$ (shown for $U = 12t$ and $5t$) and therefore the behaviour is close to the mean field result for the Heisenberg model. Deviations of $\chi_s$ from the CW result appear at high $T > 1.5t$ due to increased double occupancy and decreased local moment, while at low $T < 0.5t$ deviations occur for $U > U_c$ due to increased longer range correlations. The strong decrease of $\chi_s$ at $T < 0.1t$ and $U > U_c$ could also be due to strong increase of spin correlations or the opening of a gap to triplet excitations. $\chi_s$ is close to the CW result even in the metallic regime for $U \lesssim U_c$ and it does not show the weak $T$ dependence characteristic of a Pauli susceptibility found in Fermi liquids, consistent with our estimate $T_{\text{coh}} < 0.1t$. These results suggest well formed local moments with short-range antiferromagnetic correlations and bad-metallic behavior. Unlike the charge susceptibility $\chi_c$ the spin susceptibility $\chi_s$ does not show any signature of the metal-insulator transition.

$\chi_s$ is dominated by the quasiparticle contribution and shows Pauli paramagnetic behavior.

Frustration strongly increases the density of low-lying spin excitations, which results in a significant increase of $C_V$ and $s$ at low $T$ (see Fig. S1 in [10]). $\chi_s$ is also increased at low $T$ due to suppression of longer range correlations [10]. Such an increase is a hallmark of magnetic frustration [36].

In conclusion, we have considered the thermodynamic properties of a Mott MIT, which can be driven either by interactions ($U/t$) or geometric frustration ($t'/t$). We have shown that the metallic phase near the MIT is characterized by a small charge susceptibility, large quasiparticle renormalization, a reduced coherence temperature $T_{\text{coh}} \sim 0.1t$, large entropy, and large spin susceptibility. This is in agreement with experiments on organic charge transfer salts [4]. We have argued that the large entropy is due to a large local fluctuating magnetic moment, which leads to bad metallic behaviour above $T_{\text{coh}}$. Furthermore, we have shown how frustration increases the density of low energy spin excitations and reduces the range of antiferromagnetic spin correlations in the insulating phase.

We acknowledge helpful discussions with R.R.P. Singh, B.J. Powell, J. Bonča, A.R. Wright and especially with P. Prelovšek, whom we also thank for providing some of the codes we used. We thank the J. Stefan Institute, where most of the computations were performed. This work was supported by an ARC Discovery Project grant (Project No. DP1094395).

# Supplementary Material for "Thermodynamics of a bad metal–Mott insulator transition in the presence of frustration"

J. Kokalj* and Ross H. McKenzie

*School of Mathematics and Physics, University of Queensland, Brisbane, 4072 Queensland, Australia*

(Dated: April 26, 2013)

## EFFECT OF FRUSTRATION ON THERMODYNAMIC PROPERTIES IN THE MOTT INSULATING PHASE

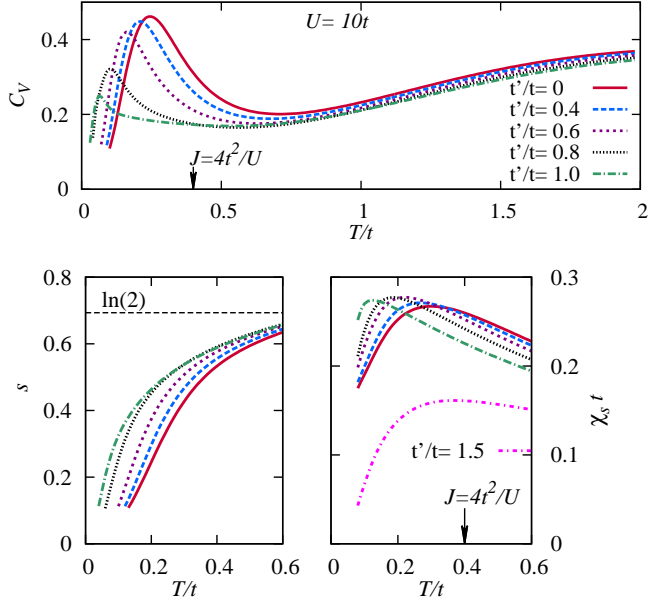

Figure S1. (color online) Effect of frustration ($t'/t$) on thermodynamic properties in the Mott insulating phase. Here we show the temperature dependence of the specific heat $C_V$ (top), entropy $s$ (bottom left) and spin susceptibility $\chi_s$ (bottom right) for $U = 10t > U_c$ and several values of $t'/t$. With increasing frustration ($t'/t$) the low energy density of states is strongly increased as seen in the strong increase of $C_v$ and $s$ at $T < 0.4t$. These states are low energy spin excitations which can be seen in increasing of $s$ towards $\ln(2)$ (free spin result) and increased low-$T$ $\chi_s$. Significant shift of the peak in $C_V$ and $\chi_s$ towards lower $T$ with increasing frustration can be attributed to decreased spin correlations. This and the increased density of low energy spin excitations is a hallmark of magnetic frustration [1]. Increasing $t'/t$ beyond 1 starts to reduce the frustration and the peak in $\chi_s$ again moves to higher $T$ (shown for $t'/t = 1.5$).

In Fig. S1 we show how frustration affects the low-$T$ properties of the insulating phase ($U > U_c$). Frustration strongly increases the density of low-lying spin excitations, which results in a dramatic increase of $C_V$ and $s$ at low $T$. Furthermore, $\chi_s$ is also increased at low $T$ due to suppression of longer range correlations and its peak moves to lower $T$. This is in agreement with series expansion results [1] for the corresponding Heisenberg models. Such dramatic increase of the low-$T$ $C_V$, $s$ and $\chi_s$ is a hallmark of magnetic frustration. In the bottom-right of Fig. S1 we also show, how with increasing $t'$ beyond $t$ and subsequently decreased geometric frustration, the peak in $\chi_s$ moves to higher $T$ again.

## FINITE TEMPERATURE LANCZOS METHOD

In this section we present in more detail the finite temperature Lanczos method (FTLM) [2, 3] and values of some of the parameters used. With FTLM one is essentially capable of obtaining results in the thermodynamic limit for temperatures $T$ above some limiting temperature $T_\text{fs}$, below which finite size effects become important. Due to geometrical frustration on the triangular lattice, the spin correlation length is substantially reduced [1], making finite size effects and $T_\text{fs}$ smaller. Additionally, we apply averaging over twisted boundary conditions [4, 5], which further reduces finite size effects, and allows exact results to be obtained in the thermodynamic limit for $U = 0$ [5]. Our results are obtained with a number of samples over different twisted boundary conditions $N_\theta \gtrsim 32$, while the effect of averaging over random vectors used in the FTLM plays only a minor role. This is due to the large number of many-body states and finite $T$, and therefore usually one random vector suffices [2, 5].

In the following we focus on the accuracy of the method in more detail, for which several parameters need to be considered. These are the number of Lanczos states $M$, the number of samples over starting random vectors $R$, and the system size $N$.

The number of Lanczos states $M$ used in our calculation varied from 50 to 100, which is orders of magnitude smaller than the number of basis states of a 16 site cluster (in which the Lanczos states are written). Such $M$s are sufficient for obtaining convergence of the result. The ground state converges within such $M$ as well as finite $T$ properties. This can be traced back to the fact that moments up to the order of $M$ are exact for a state of interest (see eq. 3.18 and corresponding text in Ref. [2]). Thermodynamics is only weakly dependent on $M$ as can be seen in top Fig. 3 in Ref. [2], where already $M = 5$ and 20 gave quite accurate results. When dealing with dynamics and spectral properties, $M$ limits the frequency resolution and usually larger values of $M$ are employed (Fig. 4 and 5 in Ref. [2]).

In contrast to the zero $T$ Lanczos method, one employs within FTLM averaging over random vectors in order to calculate finite $T$ properties. This is most nicely described and justified in section 3.5 in Ref. [2]. It is shown that the relative

statistical error is of the order of

$$\frac{\delta X}{X} \sim \mathcal{O}(\frac{1}{\sqrt{RZ}}), \qquad (S1)$$

where $R$ is the number of random vectors used and

$$Z = \text{Tr} \exp(-\beta(H - E_0)). \qquad (S2)$$

$Z$ is the thermodynamic sum normalized by $\exp(-\beta E_0)$, where $E_0$ is the ground state energy. Therefore the error is very small for large $Z$, which can appear either at high $T$ or for larger systems. Using larger systems with larger $Z$ reduces the error as well as finite $T$, since $Z$ can strongly (e.g. exponentially) increase with increasing $T$. Errors usually become larger at low $T$, where one would need to employ a large $R$ to reduce the error. In our case averaging over random vectors was in combination with averaging over twisted boundary conditions with 32 or more samples. Also, as shown in Fig. S1, frustration increases the density of states and entropy at low $T$ and helps improve accuracy in this respect. Some dependence on $R$ can be found in the bottom panel of Fig. 3 in Ref. [2], where a system with smaller Hilbert space (and therefore smaller $Z$) was used. Therefore the most computationally demanding regime within FTLM is low $T$, which can be made less demanding by improvements suggested with the low temperature Lanczos method (LTLM) [6].

More challenging are the finite size effects, which are largest at $T = 0$ (potentially long correlation lengths), but become smaller with increasing $T$, since correlation lengths decrease with increasing $T$. In Ref. [2] (section 3.7) it is argued, that finite size effects are small for $T$ above $T_{\text{fs}}$, at which $Z$ reaches a certain value ($\sim 30$). At such an elevated $T$, system size ($N$) dependence of the results becomes small and one essentially obtains a result close to the result for an infinite system. We show some system size dependence in Figs. S2 and S3. Frustration (present in our model) reduces the spin correlation length and also reduces the relevant energy scales (for example see Fig. S1), which makes $T_{\text{fs}}$ smaller. Further reduction of the finite size effects can be obtained with averaging over twisted boundary conditions [4, 5] (Fig. S2).

To shortly summarize the above discussion one can use the approximation for (S2)

$$Z(T) \sim \exp[Ns(T)], \qquad (S3)$$

where $N$ is the number of sites and $s(T)$ is the entropy per site. From this we see, that at fixed (low) temperature $s(T)$ is significantly increased by the interactions (Fig. 4) and the frustration (Fig. S1) and so this (i) reduces finite size effects, (ii) extends to lower temperatures the regime of reliability of the numerical method, and (iii) reduces the statistical noise. Thus, the numerical method is most reliable in the parameter regime of greatest physical interest: strong interactions and large frustration.

## TESTING THE NUMERICAL METHOD

In Fig. S2 we show how averaging over twisted boundary conditions improves results and makes the finite size effects smaller. In Fig. S3 we show the system size dependence of the specific heat $C_V$.

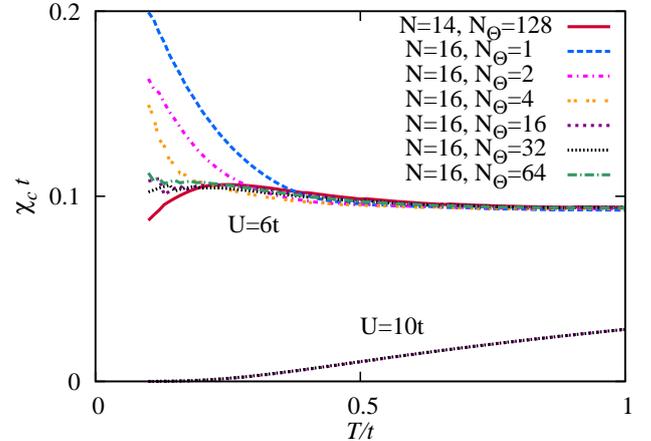

Figure S2. (color online) Averaging over twisted boundary conditions reduces finite size effects. The temperature dependence of the charge susceptibility is shown for different numbers of samples over boundary conditions $N_\theta$ for $U = 6t$ and $10t$. For $U = 6t$ and $N_\theta = 1$ and system size $N = 16$, $\chi_s$ shows a strong increase for $T < 0.4t$, which is a finite size effect. For example, $N = 14$ results for large $N_\theta$ shows no such increase and actually shows suppression for $T < 0.2t$. With increasing $N_\theta$, $\chi_c$ for $N = 16$ no longer shows strong increase at low $T$ and becomes only weakly $T$ dependent as expected for a metal and is therefore closer to the result in the thermodynamic limit. Changes of $\chi_c$ with larger $N_\theta$ are small in the insulating phase ($U = 10t$), which can be traced back to the vanishing of the Drude weight, since the Drude weight can be calculated with the derivative of the energy with respect to the twisted boundary phase $\theta$ [7–9].

## DEPENDENCE ON CLUSTER SHAPE

Within the FTLM several clusters with the same size $N$ but different shapes can be used. Different cluster shapes may give different results and one should chose a cluster that would give results with the least finite size effect and consequently closest to the thermodynamic limit. Several criteria can be considered. These are the possible frustration of the expected underlying order (standard AFM or 120 degree order), imperfection, and symmetry of the cluster. Imperfection of the cluster measures deviations from the best possible configuration of the number of independent sites in each $k$-th nearest neighbor shell [10, 11]. The best configuration would have all the nearest neighbour shells up to some shell $k$ complete (with the same number of sites in the shell as for the infinite lattice), shell $k$ would be incomplete, and all the higher shells would be empty. The imperfection measures absolute devia-tions from such a configuration. In Ref. [11] it was shown that the most important criteria is the cluster's imperfection and that numerical results from clusters with the lowest imperfection scale much better with the system size than the systems with larger imperfection. We followed these guidelines and used systems with the smallest possible imperfection (see Fig S4).

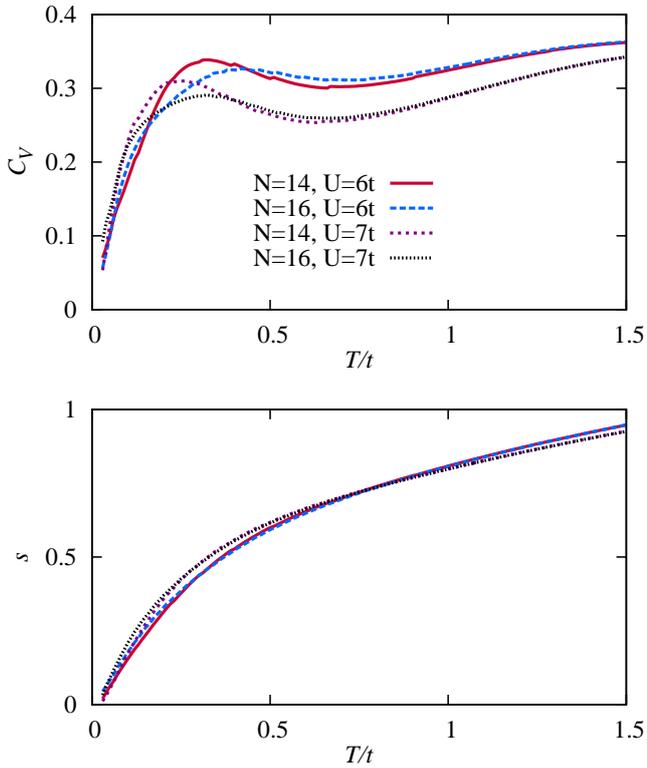

Figure S3. (color online) System size dependence of the specific heat $C_V$ and entropy $s$. For systems with $N \sim 16$, $C_V$ shows only small quantitative changes with system size at intermediate $T(\sim 0.3t)$, while at high- and low-$T$ shows only weak dependence on the system size.

Figs. S5 and S6 show the effect of different cluster shapes for a 12-site system. One can see that although the cluster shape has some effect at the intermediate temperatures, the shape of the clusters does not have any significant effect at lower and higher temperatures. For example, the linear in $T$ slope of the specific heat below the coherence temperature shows a very weak dependence on the cluster shape and size (see Fig. S6). The effect of the cluster shape in our results is also reduced due to the averaging over twisted boundary conditions and finite temperature.

## FURTHER DISCUSSION OF $\chi_c$

At zero temperature and for $U = 0$, $\chi_c = 2N_0(\mu)$, where $N_0(\mu)$ is the non-interacting electron density of states. In Fig. 1, $\chi_s$ for the non-interacting electron case ($U = 0$) shows a peak at $T \sim 0.4t$ due to the van Hove singularity.

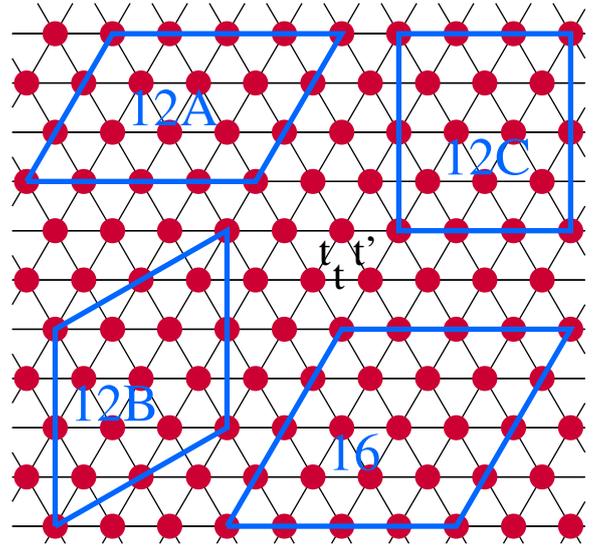

Figure S4. (color online) Different clusters used to check the dependence of results on cluster shape and size. Cluster 12A has the lowest possible imperfection (for $N = 12$) of 1, while clusters 12B and 12C have imperfection of 2. Cluster 12B does not frustrate 120 degree spin order and in addition has the $C_6$ symmetry of the infinite triangular lattice, unlike the other clusters.

$\chi_c$ decreases with increasing $U$, which is most apparent at high $T (\sim 2t)$ due to spreading of the density of states over the larger energy interval of $W + U$ or i.e. over a larger chemical potential interval. Here $W$ is a bare band width. The chemical potential interval $[\mu(n = 0), \mu(n = 2)]$ in which density of electrons changes from $n = 0$ to $n = 2$ can be determined exactly. $\mu(n = 0) = E_0(1) - E_0(0) = \epsilon_k^{min}$, where $E_0(N_e)$ is the ground state energy of a system with $N_e$ electrons and $\epsilon_k^{min}$ is the energy at the minimum of the bare band. Similarly, $\mu(n = 2) = E_0(2N) - E_0(2N - 1)$, where $N$ is the system size, $E_0(2N) = NU$ and $E_0(2N - 1) = NU - \epsilon_k^{max} - U$. $E_0(2N - 1)$ can be easily calculated with the particle-hole transformation $c_{i,\sigma} \leftrightarrow c_{i,\sigma}^\dagger$, resulting in the Hamiltonian with $t_{ij} \to -t_{ij}$ and an extra term $U(N - N_e)$ [12] and for $E_0(2N - 1)$ only one particle state needs to be considered. $\mu(n = 2) = \epsilon_k^{max} + U$. Therefore the interval of the chemical potential in which $n$ rises from 0 to 2 is increased by $U$ to $W + U$, resulting in on average decreased $\chi_c = \frac{\partial n}{\partial \mu}$. At low $T$ and metallic regime, $\chi_c$ can be related to the quasiparticle weight $z$, which is discussed in the next section, while at high $T$, $\chi_c(T \gg W) = 1/(2T)$.

In Fig. S7 we show $\ln(\chi_c)$ vs. $1/T$, which makes the opening of the charge gap $\Delta_c$ in the insulating phase clearly seen. Whether linearity of $\Delta_c(U)$ extends all the way to $U_c$ or $\Delta_c = 0$ cannot be concluded from our FTLM results, since small gaps affecting low temperatures cannot be reliably extracted due to the finite size effects. Our result of the opening of $\Delta_c$ linearly with increasing $U$ is also consistent with $T = 0$ exact diagonalization results in Ref. 13. Although their values of $\Delta_c$ are larger than ours by approximately $0.3t$, they show a trend towards our values with increasing system size. An approximately linear in $U$ opening of the gap was also found for a Hubbard model on a square lattice with next nearest neighbor hopping $t'/t = 0.2$ using the grand-canonical path-integral renormalization group [14]. Our gap values can be compared with, and appear to be smaller than, those obtained with CDMFT [15, 16]. For example we obtain $\Delta_c \sim 0.8t$ for $U \sim 9.6t$ and $t' = t$ (see Fig. 2), which is smaller than the estimate $\Delta_c \sim 1.2t$ from cellular DMFT (Fig. 5 in [15]).

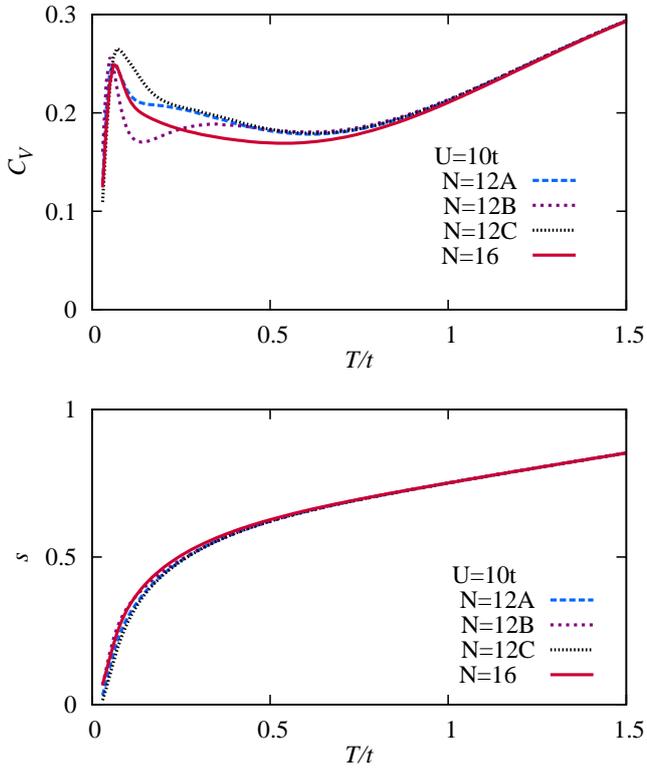

Figure S5. (color online) Cluster shape dependence of the specific heat $C_V$ and entropy $s$ in the insulating phase. Cluster size and shape does not have any significant effect on the low-$T$ specific heat and entropy. The low-$T$ peak in $C_V$ appears for all cluster shapes and results for 12A with smallest imperfection are closest to the results of the larger $N = 16$ system. All the results are for $t' = t$ and $U = 10t$.

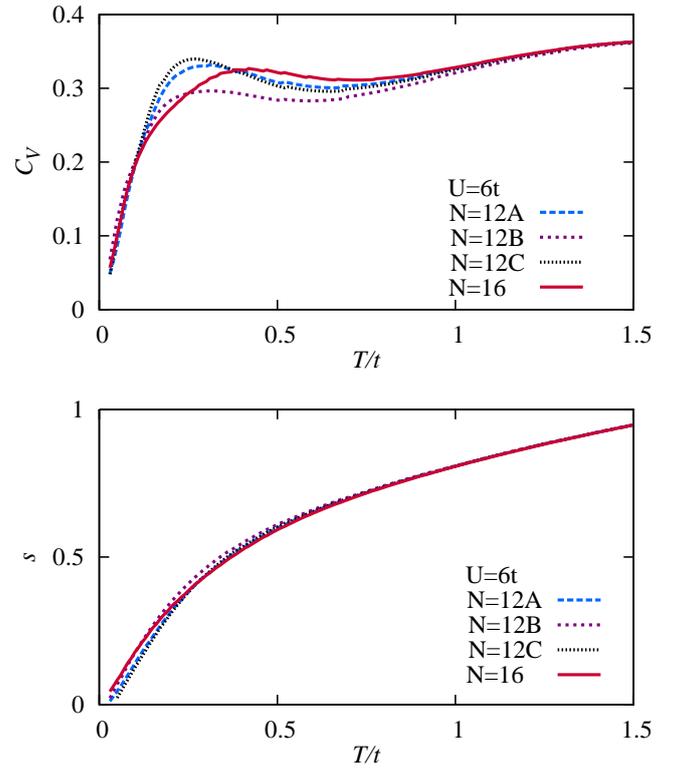

Figure S6. (color online) System size and cluster shape dependence of the specific heat $C_V$ and entropy $s$ in the metallic phase. Similar as Fig. S6 but for $U = 6t$. Cluster shape and size do not have a significant effect on entropy and specific heat below the coherence temperature.

## QUASIPARTICLE RENORMALIZATION AND CHARGE SUSCEPTIBILITY

Here we consider the effect of a quasiparticle renormalization on the charge susceptibility within a simple model and show that for a simple Fermi liquid, $\chi_c$ is proportional to a quasiparticle weight $z$. This is in contrast to what is naively expected from the increased low energy density of states due to increased renormalization or quasiparticle energies ($\epsilon_k \to z\epsilon_k$), e.g. $\chi_c \propto 1/z$.

The density of electrons $n$ in a system with $N$ sites can be calculated as

$$n = \frac{2}{N} \sum_k n_k. \quad (S4)$$

The factor 2 is due to spin, while $n_k$ can be obtained from the spectral function or imaginary part of the retarded Green's function,

$$n_k = -\frac{1}{\pi} \int d\omega f(\omega) \mathrm{Im} \frac{1}{\omega + \mu - \epsilon_k - \Sigma_k(\omega)}. \quad (S5)$$

$f(\omega)$ is a Fermi-Dirac distribution function, $f(\omega) = 1/(e^{\beta\omega} + 1)$, $\epsilon_k$ is a bare-band dispersion, $\mu$ is a chemical potential, and $\Sigma_k(\omega)$ is a self-energy. Using the definition of the charge susceptibility $\chi_c = \frac{\partial n}{\partial \mu}$ one can write

$$\chi_c = \frac{2}{N} \sum_k \frac{\partial n_k}{\partial \mu}. \quad (S6)$$

Furthermore, $\frac{\partial n_k}{\partial \mu}$ can be expressed in terms of the real $\Sigma'_k(\omega)$ and imaginary $\Sigma''_k(\omega)$ parts of the self-energy,

$$\frac{\partial n_k}{\partial \mu} = \frac{2}{\pi} \int d\omega \frac{f(\omega)\Sigma''_k(\omega)(\omega + \mu - \epsilon_k - \Sigma'_k(\omega))}{[(\omega + \mu - \epsilon_k - \Sigma'_k(\omega))^2 + (\Sigma''_k(\omega))^2]^2}. \quad (S7)$$

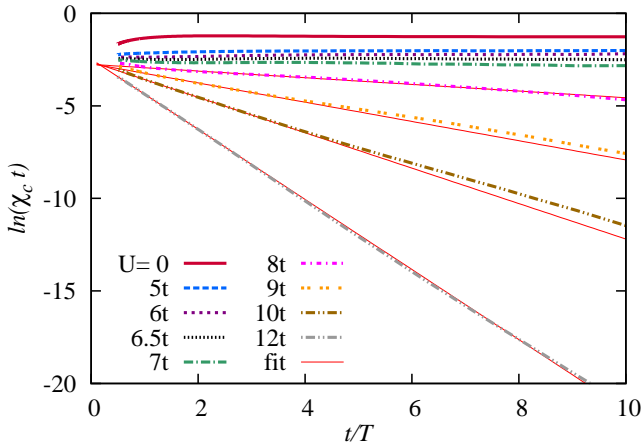

Figure S7. (color online) Arrhenius plot of $\ln(\chi_c)$ vs. $1/T$ for several interaction strengths $U$ and $t' = t$. In the insulating phase the activated behaviour is linearized in this plot, $\ln(\chi_c) = \ln(a) - \Delta_c \frac{1}{T}$, with the slope of the curves given by $-\Delta_c$. This makes the opening of the charge gap $\Delta_c$ nicely seen as increased negative slope of the curves.

Here we have assumed that the self-energy has no dependence on $\mu$. Since we are interested mostly on effects of renormalization on $\chi_c$ within the quasiparticle picture we also use the following approximations. We approximate $\Sigma'_k(\omega) = \partial_\omega \Sigma'(\omega)|_{\omega=0} \omega$, which leads to

$$\omega + \mu - \epsilon_k - \Sigma'(\omega) \sim \frac{\omega}{z} + \mu - \epsilon_k, \qquad (S8)$$

where $z \equiv 1/(1 - \partial_\omega \Sigma'(\omega)|_{\omega=0})$ and assume $\Sigma''_k(\omega) \sim \Sigma''$ or, with other words, we neglect the $\omega$ and $k$ dependence of $\Sigma''_k(\omega)$ close to the quasiparticle peak. With this approximations we can perform the integral in Eq. (S7) in the $T \to 0$ limit.

$$\frac{\partial n_k}{\partial \mu}(T \to 0) = z \frac{1}{\pi} \frac{-\Sigma''}{(\epsilon_k - \mu)^2 + \Sigma''^2}. \qquad (S9)$$

In a quasiparticle picture we assume $|\Sigma''| \ll |\epsilon_k - \mu|$, which is satisfied in the most relevant regime close to the Fermi surface and leads to the approximation

$$\frac{\partial n_k}{\partial \mu}(T \to 0) \sim z \delta(\epsilon_k - \mu). \qquad (S10)$$

The effect of quasiparticle renormalization on $\chi_c$ now becomes clear,

$$\chi_c(T \to 0) = \frac{2}{N} \sum_k z \delta(\epsilon_k - \mu) \sim z \chi_c^0, \qquad (S11)$$

where $\chi_c^0 = 2N_0(\mu)$ is the bare charge susceptibility. Charge susceptibility is due to quasiparticle renormalization reduced from it non-interacting value by a quasiparticle weight $z$.

This simple model shows that due to interactions and quasiparticle renormalization $\chi_c$ is reduced by a factor $z$, which is what we qualitatively observe in our numerical results (Figs. 1 and 2 in the main text) where we assume that at low temperatures $C_V \sim T/z$. This is in agreement with DMFT results [17], but conflicts with a claim in Ref. [18], where they suggest that both $\chi_c$ and the specific heat coefficient $\gamma$ should be proportional to $1/z$. This simple model and our results are consistent with $\chi_c \propto z$ and $\gamma \propto 1/z$.

The divergence of $\chi_c$ with approaching a Mott insulator by reduced filling [18–20] can not be captured with this simple model, and might be due to strong dependence of the self-energy on $\mu$ and/or breakdown of a quasiparticle picture.

## FURTHER DISCUSSION OF THE PHASE DIAGRAM

Deviations from perfect nesting and $t'/t = 0$ results in a strong exponential or superlinear increase of $U_c$. In Ref. [21], this was observed by shifting the chemical potential and the dependence of the critical chemical potential $\mu_c$ on $U$ is given by $\mu_c \sim \sqrt{tU} \exp(-2\pi\sqrt{t/U})$. Furthermore, strong increase of $U_c$ with increasing next nearest neighbor hopping in the square lattice was also observed within a Hartree-Fock approach [22] and a similar result is obtained for small $t'/t$ on the anisotropic triangular lattice as is shown in Fig. 3 with black dotted line.

Describing the behaviour of $U_c(t')$ at small $t'$ seems more challenging, since Hartree-Fock approximation gives a superlinear increase of $U_c$ with $t'$ [22], exact diagonalization [23] and VMC [20] suggest a linear increase, while cellular DMFT [24] result seems to be more consistent with quadratic-$t'$ dependence.

In Table I we compare critical values of $U_c$ for MIT as obtained by different methods.

The large difference between exact diagonalization results from Ref. [23] ($U_c \sim 7t$ for $N = 16$) and Ref. [25] ($U_c \sim 12t$ for $N = 12$) needs additional comment since it misleadingly suggests large finite size effects. In Ref. [23] a 16 site cluster was used without the application of twisted boundary conditions (TBC). The MIT was determined from the largest slope of the bond order, double occupancy and spin structure factor. These variables do not give an unambiguous determination of a MIT since their variation could denote transitions between different metallic phases. We instead use the charge susceptibility and charge gap, which give a direct indication of a MIT. We use averaging over TBC, finite temperature, and different criteria for determination of MIT and we obtain an estimate of $U_c/t = 7.5 \pm 0.5$, which is comparable to an estimate based on Fig. 4 in Ref. [23]. In Ref. [25] the authors used a 12 site cluster, applied just one TBC phase to obtain a closed shell configuration, and identified the MIT from a discontinuity in the Drude weight. These different boundary conditions and different criteria for MIT than the ones we use led to the estimate $U_c \sim 12t$, significantly larger than our estimate. Such a large difference comes from the particular choice of phase for the TBC used in Ref. [25]. In contrast, our estimates of $U_c$ are $7.2 \pm 0.6$ for $N = 12$ and $7.1 \pm 0.7$ for $N = 14$ and therefore show much smaller finite size effects (Compare Figure S8). Although we cannot precisely determine $U_c$ at $T = 0$ from finite $T$ calculations, our results give strong support that finite size effects are not important.

In the main text we suggest that there might be a transition at $U \sim 10t$ since a sharp low-$T$ peak appears in $C_v$ for $U \geq 10t$. Such a transition could possibly be from a spin liquid to Néel ordered phase as suggested in [26]. Determination of such a transition and characterization of the different insulating phases should be done with the calculation of additional quantities such as the spin structure factor. We merely point out that the emergence of a sharp low-T peak does not appear to be a finite size effect. For $U \geq 10t$ it appears also for smaller systems as shown in Fig. S5 and we do not observe it for $U = 8t, 9t$ for any considered size ($N = 12, 14, 16$).

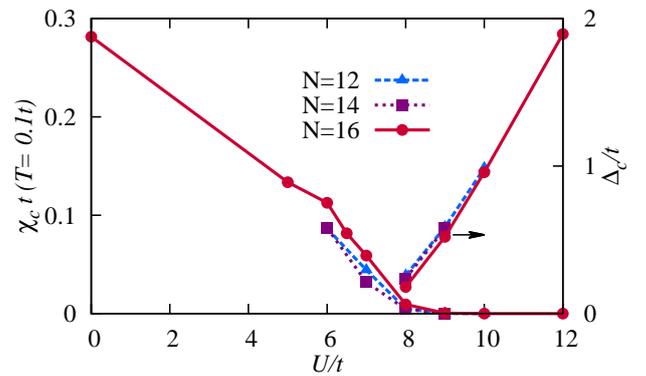

Figure S8. (color online) Weak finite size dependence of $U_c/t$ for the metal-insulator transition. The left scale shows the low temperature charge compressibility versus $U$. The right scale shows the charge gap versus $U$. In the proximity of the metal-insulator transition there is little variation in the results between lattices of $N = 12, 14, 16$ sites. All results are for $t' = t$.

Table I. Comparison of the critical interaction strength $U_c$ for a MIT at $t' = t$ as estimated by different methods. DMFT denotes dynamical mean-field theory.

| Method | $U_c/t$ | Reference |
|---|---|---|
| Slave rotors | 5.14 | [27] |
| Path-integral renormalization group | 5.2 | [28] |
| Hartree-Fock | 5.27 | [29] |
| Strong coupling expansion | 6.7 | [26] |
| Variational cluster approximation | 6.7 | [30] |
| Exact diagonalization for $T = 0$ ($N = 16$) | 7 | [23] |
| Slave boson with magnetic order | 7.23, 7.68 | [25, 31] |
| FTLM | 7.5±0.5 | this work |
| Variational Monte Carlo | 7.65±0.05 | [32] |
| Cellular DMFT | 8.5, 10.5 | [24, 33] |
| Cluster DMFT | 9.2-9.6 | [16, 34] |
| DMFT | 12-15 | [35, 36] |
| Exact diagonalization for $T = 0$ ($N = 12$) | 12 | [25] |
| Resonating-valence-bond theory | 12.4 | [37] |
| Brinkman-Rice | 15.8 | [25] |

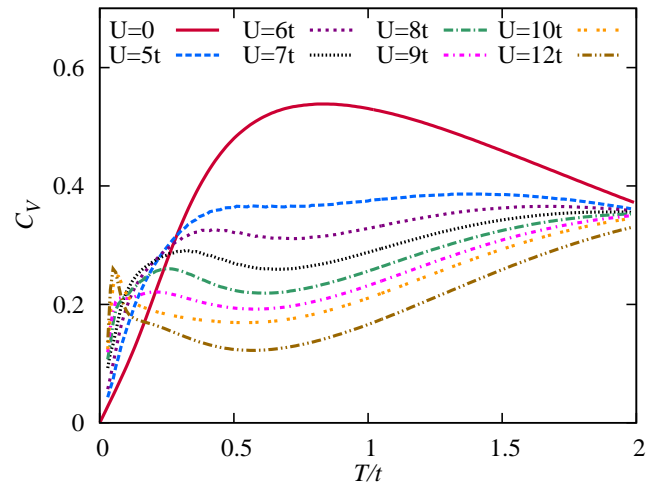

Figure S9. (color online) Specific heat $C_V$ vs. $T$ for several values of interaction $U$ and for $t' = t$. With increasing $U$, $C_V$ develops two peaks, one at low $T$, which is at large $U > U_c$ due to spin excitations, and one at high $T$ corresponding to charge excitations across the charge gap to the upper Hubbard band.

## $C_V$ AT HIGHER $T$

For $U \sim U_c$, $C_V$ already starts to develop a two peak structure (see Fig. S9), which is most pronounced for $U$ deep in the insulating regime ($U \sim 12t > U_c$). The low $T$ peak corresponds to spin excitations as was already observed for a Hubbard model on a square lattice [5], where the low-$T$ peak appeared at $T \sim J$ and for large $U$ can be captured with the Heisenberg model. In our case the peak is moved to lower $T$ ($< J$) due to frustration. The high-$T$ peak corresponds to charge excitations across the charge gap $\Delta_c$ into the upper Hubbard band.

## COMPARISON WITH EXPERIMENTAL $C_V$

Measurements have been made of the temperature dependence of the specific heat $C_V$ for a range of organic charge transfer salts [38, 39]. It is desirable to compare our calculations with these measurements, which are generally fit to the dependence $C_V(T) = \gamma T + \beta T^3$, where the first term is associated with a Fermi liquid (possibly of spinons) and the second with lattice vibrations. Unfortunately, the reported experimental data is mostly at temperatures below $0.03t \sim 10$ K, which is less than the temperature at which the FTLM is reliable. Nevertheless, we see some features of the experimental data that are relevant.

Even though $J = 4t^2/U \simeq t/2 \simeq 200K$ there is a significant entropy, of order $0.2\ln(2)$ at $T = 0.05t \ll J$ (see Fig. 4). A similar estimate was also made by Manna et al. [40]. Indeed, Yamashita et al. [38] estimated that the spin entropy of $\kappa$-(BEDT-TTF)$_2$(CN)$_3$ was of order $(0.1 - 0.2)\ln(2)$ at 10 K. They suggested that "*this entropy release is unexpectedly large for an antiferromagnet with $J \sim 250K$ and provides additional evidence for the realization of a spin liquid with large degeneracy*". Our result in Fig. 4 suggests that the released entropy can be large also when the ground state is antiferromagnetically ordered (e.g., result for $U \gtrsim 12t$). The peak they observe around 6 K is beyond the temperature range of our study. We also cannot reliably extract the linear term $\gamma$, since it would again require very low-$T$ results.

One should however be cautious when extracting the electronic contribution from the measured $C_V$ due to the large lattice contribution. Although varying the counterion $X$ in $\kappa$-(BEDT-TTF)$_2$X and in $\beta'$-X[Pd(dmit)$_2$]$_2$ does change the electronic contribution at low $T$ in the $C_V$ due to the changes in $U/t$ and $t'/t$ (see Fig. 4), it also changes the lattice contribution. For example, typical variations of $\beta$ with X are $\sim 10$ mJ/K mol [38, 39], which already at $T \sim 20$ K lead to the difference in the released entropy of $\sim 3R$. This is much larger that the maximal possible release of the entropy from spins $\sim \ln(2)R$. This together with the right trend (larger $\beta$ for a smaller Debye frequency associated with a heavier counterion X, e.g., in Fig. 2 in Ref. [39] ) suggest that $\beta$ is dominated by the lattice vibrations, while the smaller spin contribution to $C_V$ is hard to extract.

## DOUBLE OCCUPANCY AND LOCAL MAGNETIC MOMENT

The double occupancy $D$ can be calculated with the use of the free energy $F$ defined by

$$e^{-\beta F} = \text{Tr}\left[e^{-\beta H}\right]. \quad \text{(S12)}$$

Usually one evaluates $D$ by taking the derivative of $F$ with respect to $U$ at fixed chemical potential $\mu$. On the other hand, we are dealing with a fixed number of particles or fixed filling and therefore in our case $\mu$ changes with $U$. Taking this into account, one gets

$$D = \langle n_{i,\uparrow} n_{i,\downarrow} \rangle = \frac{1}{N}\left(\frac{\partial F}{\partial U}\bigg|_{N_e} + N_e \frac{\partial \mu}{\partial U}\bigg|_{N_e}\right). \quad \text{(S13)}$$

Therefore, $D$ is calculated from the derivative of $F$ with respect to $U$ at fixed number of electrons $N_e$, and one needs to add a term due to the change of chemical potential with $U$ at fixed $N_e$. $N$ is a number of sites in the system. Our calculation of $D$ serves only as a rough estimate, since we take numerical derivatives of $F$ and $\mu$ for quite large $\Delta U \sim t$. This does not allow for a precise determination of $D$, and smooths out any sharp features of $D$ as a function of $U$.

In Fig. S11 we show the calculated $U$ dependence of a local moment $\langle s_z^2 \rangle$, which shows a smooth behaviour without any substantial change near $U_c$. This supports the picture of a large local moment, with values close to the strong coupling (Heisenberg) limit even in the metallic phase. Our results do not show the behavior of a local moment predicted with the Hartree-Fock or Slater approximation, where the local moment is a constant with the non-interacting value for $U < U_c$, and increases slowly with increasing $U$ for $U > U_c$. In this approximation the MIT is driven by antiferromagnetism, while our results are consistent with the MIT driven by Mott physics. Our results are also in contrast with the Brinkman-Rice picture [41], which predicts that the local moment increases linearly with increasing $U$ for $U < U_c \sim 15.8t$, fully develops ($4\langle s_z^2 \rangle = 1$) at $U = U_c$ and stays constant for $U > U_c$.

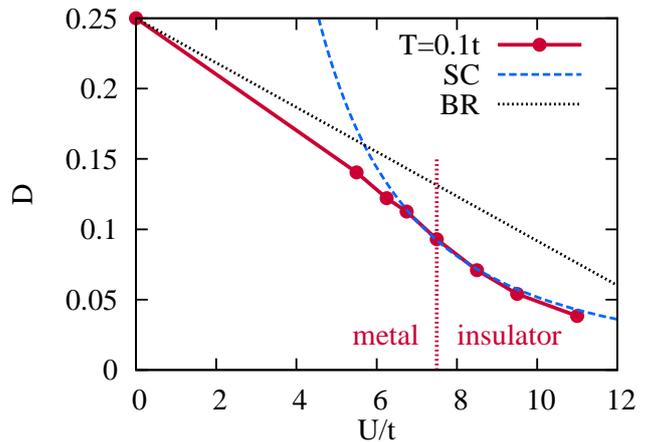

Figure S10. (color online) Double occupancy $D$ vs. interaction strength $U$ for $t' = t$ and $T = 0.1t$. $D$ decreases with increasing $U$ as expected. For small $U < U_c$, $D$ shows linear decrease with increasing $U$, which is predicted with the Brinkman-Rice (BR) picture [41] ($D^{(BR)} = (1 - U/U_c^{(BR)})/4$). However, the BR picture predicts too large a $U_c^{(BR)} \sim 15.8t$ [25, 41]. Close to $U_c$, $D$ is more strongly suppressed and starts to exhibit strong coupling (SC) [12, 42] or Heisenberg behaviour in which double occupancy is given by [42] $D^{(SC)} = (2t^2/U^2) \sum_\delta (1/4 - \langle \mathbf{S}_i \cdot \mathbf{S}_{i+\delta} \rangle)$. The sum over $\delta$ goes over all 6 nearest neighbours and the strong coupling result is shown for $\langle \mathbf{S}_i \cdot \mathbf{S}_{i+\delta} \rangle \sim -0.182$ [43]. This estimate of spin correlation is evaluated within Heisenberg model and is valid for $U > U_c$, where it shows only a weak dependence on $t/U$ [12] . The agreement of the calculated $D$ and the SC result is surprisingly good in the regime shown in the figure ($U \gtrsim U_c$). The small value of $D \sim 0.1$ close to $U_c$ corresponds to only every tenth site being doubly occupied, which results in a large local moment and strong spin response manifested in large $\chi_s$. We note that in Ref. 26 a small discontinuity ($\sim 0.01$) in $D$ was observed at $U \simeq 10t$ and attributed to a first-order transition from a spin liquid ($6t < U < 10t$) to a Néel antiferromagnet with 120 degree spiral order ($U > 10t$). Our results do not have sufficient resolution to detect such a transition.

The agreement with the strong coupling result seems surprisingly good for $U$ close to $U_c$, which suggests that the Heisenberg model gives a good approximation also in the regime $U \gtrsim U_c$ and that higher order terms do not play a crucial role. This appears in contrast with results in Ref. 26, where they observed that higher order terms are actually responsible for the transition between Néel ordered and spin liquid phase. However, our results are for finite $T$ where the small differences in the ground state energies are not that important, and also the change of $D$ at the transition was observed to be only a few percent [26]. Furthermore, we estimate $D$ by numerically differentiating the free energy over $U$ with $\Delta U \sim t$, which further smooths the $U$-dependence of $D$.

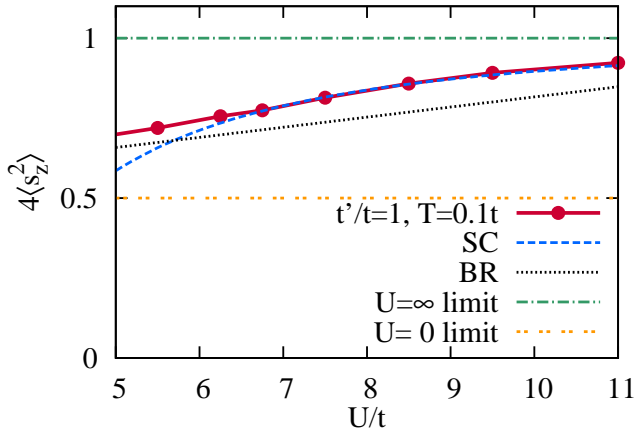

Figure S11. (color online) Local moment $4\langle s_z^2 \rangle = 1 - 2D$ vs. interaction strength $U$ for $t'/t = 1$ and $T = 0.1t$. $4\langle s_z^2 \rangle$ is increasing with increasing $U$ as expected and has values close to the ones expected from the strong coupling (SC) limit [42] for $U > U_c$. Large values of the local moment persist also in the metallic regime for $U < U_c$ and we do not observe a strong decrease of $\langle s_z^2 \rangle$ with decreasing $U$ at the MIT. Therefore the metallic phase is characterized also with a large local moment and strong spin response, e.g., with large $\chi_s$. The Brinkman-Rice result [41] is also shown, together with limiting values for $U = \infty$ and $U = 0$.

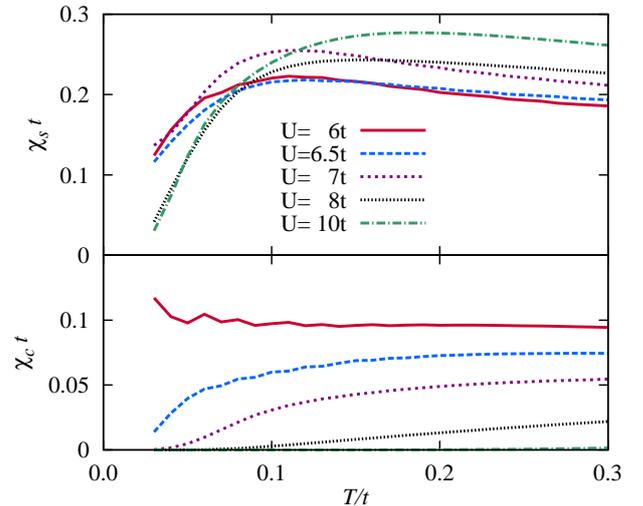

Figure S12. (color online) Sign of a pseudogap in the metallic phase close to the Mott MIT. The temperature dependence of the spin ($\chi_s$) and charge susceptibilities ($\chi_c$) are shown for $t'/t = 0.8$. The suppression of $\chi_s$ for $U = 6t$ and $0.03t < T < 0.1t$ (top panel) could be due to the emergence of a pseudogap. In this regime the system is metallic, as can be seen in the $T$ independent $\chi_c$ (bottom panel). This is in agreement with the measured Knight shift $K_s$ [44]. However, caution is in order, since the suppression of $\chi_s$ for $U = 10t$ at $T < 0.05t$ may be over-estimated due to finite size effects (see text).

## PSEUDOGAP

An important question is whether a pseudogap is present in the metallic phase close to the Mott insulator [45]. Signatures of such a pseudogap are seen in NMR experiments on $\kappa$-(BEDT-TTF)$_2$Cu[N(CN)$_2$]Br [44, 46]. The Knight shift, which is proportional to the spin susceptibility $\chi_s$, decreases by about 40% as the temperature is lowered from about 50 K to 10 K. The NMR relaxation rate $1/T_1T$ increases with decreasing temperature, with a maximum around 50 K, and then decreases by about a factor of two as the temperature is lowered to 10 K. These temperature dependences are qualitatively similar to what is observed in underdoped cuprate superconductors for which ARPES provides independent evidence of a pseudogap which is anisotropic in momemtum space.

Figure S12 shows that for $t' = 0.8t$ and $U = 6t$, $\chi_s$ decreases by about 50% as the temperature decreases from about $0.1t$ to $0.03t$. These parameters correspond to the metallic phase, as indicated by the non-zero charge compressibility. The calculated temperature dependence appears to be consistent with the experiment [$t \simeq 500$ K]. However, caution is in order, because of the possible role of finite size effects at such low temperatures. This can be seen by examining the temperature dependence of $\chi_s$ for $U = 10t$ which is in the Mott insulating phase. It has a maximum around $T = 0.15t$ and decreases smoothly to zero around $0.03t$. Similar behaviour is found for $U = 10t$ and smaller values of $t'/t$. These results can be compared to known results for the corresponding Heisenberg model. In particular for $J' < 0.5J$ [$t' < 0.7t$] the model should have long-range Néel order at zero temperature. Quantum Monte Carlo simulations on the square lattice model show the temperature dependent spin susceptibility has a maximum around $T \simeq J$ and then decreases by about 50% to a non-zero value at $T = 0$; hence, there is no spin gap [47]. This discrepancy shows that the apparent gap observed in FTLM is a finite size effect. On the other hand, the suppression of $\chi_s$, in the temperature range $0.05t < T < 0.1t$ may be a real effect. But, it is not clear at a moment what is the physics behind this reduction in the spin susceptibility. It could be due to a suppression of the density of states such as associated with a pseudogap. Or like in the Heisenberg model the reduction could be due to the development of longer-range antiferromagnetic correlations in the bad metallic phase.

---